\documentclass{ws-procs975x65}
\usepackage{amsmath}
\begin{document}

%to switch ON running title
%\markboth{L. Hatcher}{Quantum States from Tangent Vectors}

\title{THE GENERALIZED SECOND LAW IN DARK ENERGY DOMINATED UNIVERSES}

\author{Germ\'{a}n Izquierdo$^*$ and  Diego Pav\'{o}n$^\dag$}

\address{Department of Physics,\\
 Autonomous University of Barcelona, \\
 08193 Bellaterra (Barcelona),\\
Spain\\
$^*$E-mail: german.izquierdo@uab.es\\
$^\dag$E-mail: diego.pavon@uab.es }

\begin{abstract}
The generalized second law of gravitational thermodynamics is
examined in scenarios where the dark energy dominates the cosmic
expansion. For quintessence and phantom fields this law is
fulfilled but it may fail when the dark energy is in the form of a
Chaplygin gas. However, if a black hole is allowed in the picture,
the law can be violated if the field is of phantom type.
\end{abstract}
\keywords      {Theoretical cosmology, black holes,
thermodynamics}

\bodymatter

\section{Introduction}
Nowadays, the overwhelming observational evidence suggests that
the Universe is undergoing an accelerated expansion driven by some
form of energy (dubbed ``dark energy") that violates the strong
energy condition, $\rho + 3p >0$. What is more, models in which
this field also violates the dominant energy condition (DEC),
$\rho+p> 0$ -dark energy of ``phantom" type- appears marginally
 favored by the data. This short Communication  briefly summarizes our
research about the validity of generalized second law (GSL) of
gravitational thermodynamics in spatially flat, dark energy
dominated, FLRW universes \cite{plb1,plb2}. This law asserts that
the entropy of matter and fields within the event horizon plus the
entropy of the horizon is a never decreasing quantity. Event
horizons, of radius $R_{H} = a(t)\int_{t}^{\infty}{dt'/a(t')}$,
are unavoidable features of ever-accelerating cosmologies and are
widely assumed to possess thermodynamic properties as temperature
and entropy, though  this has been shown in a rigurous manner for
the de Sitter horizon only \cite{gary}.

\section{The GSL in accelerated universes}
We begin by considering a phantom dominated universe with equation
of state parameter $w =$ constant $<-1$. Its horizon radius $R_{H}
= n/[(1+n)H]$, with $n= -2/[3(1+w)]$, decreases with expansion
whence its associated temperature, $T_{H} = n/[(1+n)2\pi R_{H}]$
augments with time while its entropy, $S_{H} = \pi R_{H}^{2}$,
diminishes. However, the entropy of the phantom, assumed in
thermal equilibrium with the horizon, calculated through Gibbs'
equation \cite{Herbert} exactly compensates the decrease of the
horizon entropy, $S= -S_{H}$, whereby the GSL is satisfied. (It is
noteworthy that, in agreement with previous authors
\cite{previous}, the entropy of the phantom fluid is negative.
However, strange as it may be, this is not forbidden by any
thermodynamical law). A parallel study for non-phantom dark energy
dominated cosmologies, with $-1<w =$ constant $<-1/3$
 (e.g., ``quintessence"), yields identical result only that, now,
 $S$ is positive and decreasing, and $S_{H}$ augments.

As an example of a phantom-dominated expansion with variable $w$
we consider the model of Sami and Toporensky \cite{sami} which
features a scalar field with negative kinetic energy and potential
$V = V_{0}\, \phi^{\alpha}$, where $0< \alpha \leq 4$. A similar
analysis shows that $\dot{S}+ \dot{S}_{H} \geq 0$, the equality
sign holding just for $t \rightarrow \infty$, i.e., when $R_{H}$
vanishes. Again, $S$ is negative and increasing and such that
$|S|/S_{H} \geq 1$.

The Chaplygin gas \cite{chaplygin}, of equation of state $p =
-A/\rho$, corresponds to a non-phantom dark energy field with $w =
- A a^{6}/(Aa^{6}+B)$. Here $A$ and $B$ are positive-definite
constants. In this case, the radius of the horizon increases with
expansion up to the de Sitter value, $H^{-1}$, for $t \rightarrow
\infty$, as it should -see, Fig. 3 of Ref. \cite{plb1}-, and the
GSL is fulfilled for $a \geq [2.509\, B/A]^{1/6}$ but it may be
violated at earlier times.

We next consider the impact of a small black hole within the event
horizon in phantom dominated universes on the validity of the GSL.
By ``small" we mean that the black hole mass is much lower than
the phantom energy inside the horizon, i.e., $M/E_{\phi} \ll 1$,
so that neither the scale factor nor the event horizon radius gets
significantly modified. As is well known, Schwarzschild black
holes, immersed in a phantom environment, are bound to lose mass
by accreting phantom energy at a rhythm  $\dot{M} = -16 \pi%
M^{2}\, \dot{\phi}^{2}$ -see Ref.\cite{babichev}. Therefore, the
black hole entropy, $S_{BH} = 4 \pi M^{2}$, will necessarily
decrease. Thus, for scenarios with $w =$ constant $< -1$ it
follows that $\dot{S}+ \dot{S}_{BH}+ \dot{S}_{H} < 0$, i.e., the
GSL is violated.

It remains to be seen whether it will be also violated when $w$
varies with time. To this end we consider again the model of Sami
and Toporensky \cite{sami}. In this scenario, $\dot{S}_{BH} = -8%
\pi \alpha H^{2} M^{3}\, x^{-1}$, with $x \equiv 4\pi%
\phi^{2}/\alpha$, and the GSL is satisfied provided the black hole
mass does not exceed the critical value \cite{plb2},
\\
\begin{eqnarray}
M_{cr}&=&
\sqrt{\frac{3}{8V_{0}}}(4\pi)^{-\frac{1}{2}+\,\frac{\alpha}{4}}\,
\alpha^{-\frac{1}{3}-\,\frac{\alpha}{4}}x^{\frac{1}{3}-\frac{\alpha}{12}}
\nonumber \\
&\times&\left\{e^{x}\Gamma \left(\frac{4-\alpha }{4},\,
x\right)\left[\Gamma \left( \frac{4-\alpha }{4},\, x\right)\,
e^{x} \left(2 + \frac{\alpha}{2x}\right)- 2 x^{-\alpha/4}
\right]\right\}^{\frac{1}{3}}\, , \nonumber
\end{eqnarray}
\\
which decreases with time at fixed $\alpha$  much slowly than $M$.
This implies that, regardless the initial mass of the black hole,
sooner or later we will have $M > M_{cr}$ while the condition
$M/E_{\phi} \ll 1$ still holds -see Fig. 1 in Ref. \cite{plb2}. As
a consequence the GSL will be violated. At some point further
ahead the said condition will no longer be met and our analysis
will break down. From this point on we can say nothing about the
validity of the GSL.

\section{Discussion}
The assumption of thermal equilibrium between the dark energy and
the event horizon may seem artificial. However, this condition
must be fulfilled for the entropy concept to be meaningful. In
other words, the entropy is an exclusive  property of  equilibrium
systems whence the entropy of two systems cannot be meaningfully
added unless they are into equilibrium with one another
\cite{Herbert}.

In view of the failure of the GSL in phantom dominated scenarios
when black holes are present different reactions may arise: $(i)$
Some phantom energy fields might be physical but not those
considered in this Communication. Indeed, several predictions
lending support to phantom fields may have come from an erroneous
interpretation of the observational data \cite{das}. $(ii)$ The
GSL was initially formulated for systems complying with the DEC,
so there is no reason why it ought to be satisfied by systems that
violate it. $(iii)$ Strictly speaking, a general proof of the GSL
even for systems complying with the DEC is still lacking
\cite{bobw}, therefore we should not wonder at its failure in some
particular cases.

It is for the reader to decide which of these alternatives, if
any, is more to his/her liking.

Yet, one may argue that it is unclear that black holes retain
their thermodynamic properties (entropy and temperature) in
presence of a field that does not comply with the DEC. In such an
instance, one may think, that there is no room for the black hole
entropy in the expression for the GSL. However, the latter is
often formulated by replacing $S_{BH}$ by the black hole area.
Again, this variant of the GSL will fail in the two cases of
above. To sum up, if eventually phantom energy is shown to be a
physical reality, it will pose a serious threat to the generalized
second law.

\bibliographystyle{ws-procs975x65}

\begin{thebibliography}{10}
\bibitem{plb1}
G. Izquierdo and D. Pav\'{o}n, {\it Phys. Lett. B} {\bf 633}, 420
(2006).
\bibitem{plb2}
G. Izquierdo and D. Pav\'{o}n, {\it Phys. Lett. B} {\bf 639}, 1
(2006).
\bibitem{gary}
G. Gibbons and S.W. Hawking, {\it Phys. Rev. D} {\bf 15}, 2738
(1977).
\bibitem{Herbert}
H. Callen, {\it Thermodynamics} (J. Wiley, N.Y., 1960).
\bibitem{previous}
M.D. Pollock and T.P. Singh, {\it Class. Quantum Grav.} {\bf 6},
901 (1989); I. Brevik {\em et al.}, {\it Phys. Rev. D} {\bf 70},
043520 (2004); J.A.S. Lima and J.S. Alcaniz, {\it Phys. Lett. B}
{\bf 600}, 191 (2004).
\bibitem{sami}
M. Sami and A. Toporensky, {\it Mod. Phys. Lett. A} {\bf 19}, 20
(2004).
\bibitem{chaplygin}
A. Kamenshchik, U. Moschella and V. Pasquier, {\it Phys. Lett. B}
{\bf 511}, 265 (2001).
\bibitem{babichev}
E. Babichev, V. Dokuchev and Yu. Eroshenko, {\it Phys. Rev. Lett.}
{\bf 93}, 021102 (2003).
\bibitem{das}
S. Das, P.S. Corasaniti, and J. Khoury, {\it Phys. Rev. D} {\bf
73}, 083509 (2006).
\bibitem{bobw}
R.M. Wald, {\it Quantum Field Theory in Curved Spacetime and Black
Hole Thermodynamics} (University of Chicago Press, Chicago, 1994).
\end{thebibliography}

\end{document}